\documentclass[a4paper,12pt]{article}

\usepackage{amsmath, amsthm, amsfonts, amssymb}
\usepackage{graphicx}
\theoremstyle{plain}
\newtheorem{theorem}{Theorem}[section]

\newtheorem{lemma}{Lemma}[section]

\theoremstyle{remark}
\newtheorem{remark}{Remark}[section]

\def\Om{\Omega}
\def\om{\omega}
\def\e{\varepsilon}
\def\g{\gamma}

\def\l{\lambda}
\def\p{\partial}
\def\D{\Delta}

\def\a{\alpha}

\def\t{\widetilde}

\def\Si{\Sigma}

\def\x{x'}
\begin{document}

\title{On a problem with nonperiodic frequent alternation of boundary
condition imposed on fast oscillating sets}
%\titlerunning{On a problem with nonperiodic alternation\ldots}

%\subtitle{\empty}

\date{\empty}
\author{Denis I. Borisov
\thanks{The author is supported by RFBR (02-01-00693, 03-01-06407), by
the program ''Leading Scientific School'' (NSh-1446.2003.1) and
partially supported by the program ''Universities of Russia'
(UR.04.01.010). }
}                     % Do not remove
%
%\offprints{D.I. Borisov}          % Insert a name or remove this line
%
%\address{D.I. Borisov\at

\maketitle

\begin{quote}
{\small  Bashkir State Pedagogical University, October
Revolution, St.~3a, 450000 Ufa, Russia, \\
 E-mail:
\texttt{BorisovDI@ic.bashedu.ru}, \texttt{BorisovDI@bspu.ru},
\\
URL: \texttt{http://borisovdi.narod.ru} }
\end{quote}

\begin{abstract}
We consider singular perturbed eigenvalue problem for Laplace
operator in a cylinder with frequent and nonperiodic alternation
of boundary conditions imposed on narrow strips lying in the
lateral surface. The width of strips depends on a small
parameter in a arbitrary way and may oscillate fast, moreover,
the nature of oscillation is arbitrary, too. We obtain two-sided
estimates for degree of convergences of the perturbed
eigenvalues.
\end{abstract}

\section{Introduction}
\label{intro}

This paper is devoted to the study of eigenvalue problem with
frequent alternation of boundary conditions. The main feature of
such boundary conditions is as follows. One should define the
subset of a boundary consisting of a great number of disjoint
closely-spaced parts of small measure. The boundary condition of
one type is imposed on this subset, while one of another kind is
settled on the rest part of the boundary. Papers devoted to
homogenization of the elliptic problem with frequent alternation
of boundary condition appeared in the second half of the last
century (see, for instance, \cite{LP2,LP,Ch,Dlcr,Fr,OlCh1}).
Nonlinear elliptic boundary value problems involving frequent
alternation of boundary conditions were treated in
\cite{BBC,Dv}. In \cite{BBC} they solved the problem
numerically, while in \cite{Dv} the homogenization theorems were
proved. In the papers \cite{F1,F3} initial-boundary parabolic
problem with frequent alternation of boundary condition were
considered and, in particular, homogenization theorems were
proved.

Asymptotics expansion for the solutions to the elliptic problems
with frequently alternating boundary condition are constructed
last decade. Two-{di\-men\-sio\-nal} case for periodic structure
of alternation was studied well enough (see
\cite{VMU,Asan,AA,ZhVM} and References of these works),
nonperiodic case was considered in \cite{DAN}. Asymptotics
expansions for three-dimensional problems with periodic
structure of alternation were established in \cite{CR,BMS,BDU}.
The leading terms of asymptotics expansions for the solution to
a parabolic problem with frequent alternation of boundary
conditions were obtained in \cite{F1,F3}.

In the present paper we study eigenvalue problem for Laplace
operator in a cylinder. The lateral surface is partitioned into
great number of narrow strips on those the Dirichlet and Neumann
condition are imposed in turns. We deal with essentially
nonperiodic alternation, moreover, the width of the strips may
varies arbitrarily, including the cases of slow and fast
oscillations of complicated nature. Under minimal restrictions
to the structure of alternations we give best possible two-sided
estimates for degree of convergences of the perturbed
eigenvalues.

We also note that estimates for degrees of convergence for
another structure of alternation were obtained in
\cite{DAN,Dr,F1,F3,OlCh1}, where parts of the boundary with
different boundary condition were assumed to shrink to points.

\section{Formulation of the problem and the main results}
\label{sec:1}

Let $x=(\x,x_3)$, $\x=(x_1,x_2)$ be Cartesian coordinates, $\om$
be an arbitrary bounded simply-connected domain in
$\mathbb{R}^2$ having infinitely differentiable boundary,
$\Om=\om\times[0,H]$ be a cylinder of height $H>0$ with upper
and lower basis $\om_1$ and $\om_2$ respectively. By $s$ we
denote the natural parameter of the curve $\p\om$, while $\e$ is
a small positive parameter: $\e=H/(\pi N)$, where $N\gg1$  is a
great integer. We define a subset of the lateral surface $\Si$
of the cylinder $\Om$, consisting of  $N$ narrow strips (cf.
figure):
\begin{equation*}
\g_\e=\{x: \x\in\p\om, -\e a_j(s,\e)<x_3-\e\pi(j+1/2)<\e
b_j(s,\e), j=0,\ldots,N-1\}.
\end{equation*}
Here $a_j(s,\e)$ and $b_j(s,\e)$ are arbitrary functions
belonging to  $C^\infty(\p\om)$ and satisfying uniform on $\e$
and $j$ estimates:
\begin{equation}\label{3}
0<a_j(s,\e)<\frac{\pi}{2},\quad 0<b_j(s,\e)<\frac{\pi}{2}.
\end{equation}
From geometrical point of view these estimates means that strips
of the subset $\g_\e$ associated with different values of $j$ do
not intersect.

We study singularly perturbed eigenvalue problem:
\begin{gather}
-\D\psi_\e=\l_\e\psi_\e,\quad x\in\Om,\label{1}
\\
\psi_\e=0,\quad x\in\om_1\cup\g_\e,\qquad
\frac{\p}{\p\nu}\psi_\e=0,\quad
x\in\om_2\cup(\Si\setminus\overline{\g}_\e),\label{2}
\end{gather}
where $\nu$ is the outward normal.

\begin{figure}%[t]
\begin{center}

%\noindent
\includegraphics[height=6.36 true cm, width=5.71 true cm]
{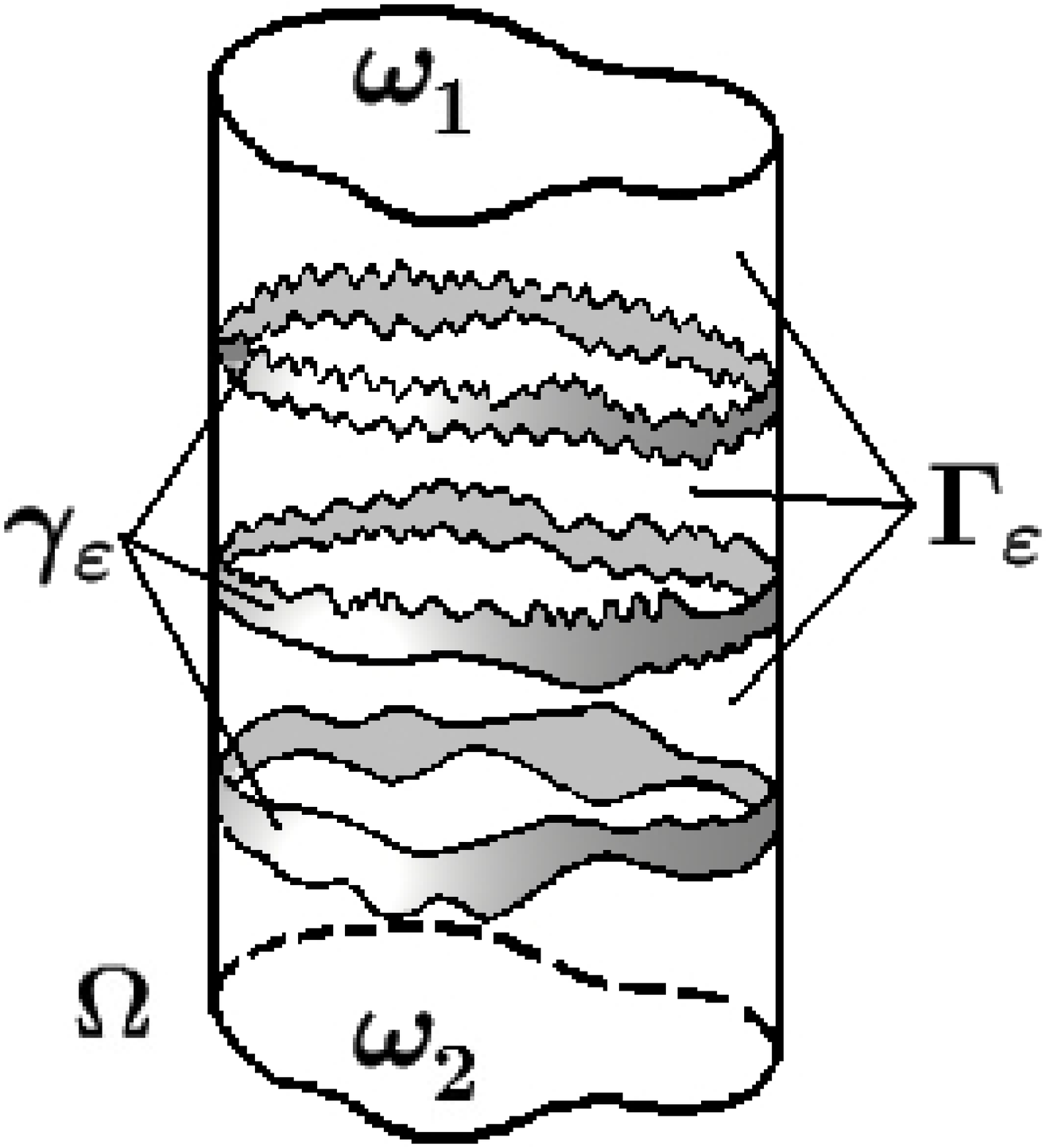}

%\medskip

%Figure.
\end{center}

\caption{}

\end{figure}

The object of the paper is to find out the limits to the
eigenvalues $\l_\e$ and to estimate the degree of convergences
as $\e\to0$ under minimal restrictions for the set $\g_\e$.
Earlier elliptic problem with boundary conditions  (\ref{2})
were studied in \cite{CR,BMS,BDU,LP}. In the paper \cite{LP}
they studied the homogenization of the elliptic problem in a
circular cylinder with boundary conditions (\ref{2}) under
additional assumption that $a_j(s,\e)=b_j(s,\e)=\eta(\e)$,
$j=0,\ldots,N-1$, where $\eta(\e)$ is a some function.
Geometrically this assumption means that the set  $\g_\e$ is
periodic and all strips have same constant width. In
\cite{BMS,BDU} under the same assumptions complete asymptotics
expansions for the problem (\ref{1}), (\ref{2}) were
constructed. The case of arbitrary section and periodic
distribution of the strips of slowly varying width was treated
in \cite{CR,BMS}. In \cite{CR,BMS} it was assumed that
$a_j(s,\e)=b_j(s,\e)=\eta(\e)g(s)$, where $g(s)\in
C^\infty(\p\om)$, $0<c_0\le g(s)\le 1$, and the function
$\eta(\e)$ satisfies the estimate $0<\eta(\e)<\pi/2$. The
leading terms of the asymptotics expansions for the
eigenelements were constructed, moreover, the definition of
$a_j$ and $b_j$ mentioned above played essential role in
construction of these asymptotics. At the same time, the
question on behaviour of the eigenvalues of the problem
(\ref{1}), (\ref{2}) for arbitrary functions $a_j$ and $b_j$ is
open. Clear, the arbitrary choice of the functions $a_j$ and
$b_j$ leads, generally saying, to the set  $\g_\e$ of
nonperiodic structure. Moreover, the functions $a_j(s,\e)$ and
$b_j(s,\e)$ may oscillate fast on $s$. Such oscillation takes
place under, for instance, following choice of the functions
$a_j$ and $b_j$: $a_j(s,\e)=b_j(s,\e)=\eta(\e)g(s/\a(\e))$,
where $\eta(\e)$ is a rough width of the strip, $g\in
C^\infty(\p\om)$, $\a(\e)\xrightarrow[\e\to0]{}0$. We stress,
that in the example shown the leading terms of the asymptotics
expansions for the eigenvalues of the problem (\ref{1}),
(\ref{2}) will differ from the similar results of the papers
\cite{CR,BMS,BDU} and will depend essentially on the function
$\a(\e)$. If the set $\g_\e$ is nonperiodic and the width of all
or some strips oscillates fast, and also, these oscillations
have arbitrary nonperiodic on $s$ structure (cf. figure), the
question of constructing the asymptotics expansions for the
eigenvalues of the problem (\ref{1}), (\ref{2}) becomes very
complicated and can be solved only under some additional
constraints for the functions $a_j$ and $b_j$. That's why to
estimate the degree of convergence for the eigenvalues of the
(\ref{1}), (\ref{2}) is a topical question. Exactly this
question is solved in the present paper.

Now we proceed to the formulation of the main results. We take
the perturbed eigenvalues in ascending order counting
multiplicity: $\l_\e^1\le\l_\e^2\le\ldots\l_\e^k\le\ldots$
Limiting eigenvalues defined below are assumed to be taken in
the same order: $\l_0^1\le\l_0^2\le\ldots\l_0^k\le\ldots$

\begin{theorem}\label{th1} Suppose the inequalities (\ref{3}) and
\begin{equation*}
\eta(\e)\le a_j(s,\e),\quad \eta(\e)\le b_j(s,\e),
\end{equation*}
hold, where $\eta(\e)$ is an arbitrary function obeying the
estimate $0<\eta(\e)<\pi/2$ and the equality:
\begin{equation*}
\lim\limits_{\e\to0}\e\ln\eta(\e)=0.
\end{equation*}
Then uniform on $\e$ and $\eta$ estimates
\begin{equation}
-c_k\e(|\ln\eta|+1)\le \l_\e^k-\l_0^k\le 0\label{7}
\end{equation}
hold true. Here $c_k>0$ are constants, $\l_0^k$ are eigenvalues
of the problem
\begin{equation}
\begin{aligned}
&-\D\psi_0=\l_0\psi_0,\quad x\in\Om,
\\
 \psi_0=0,\quad &x\in\om_1\cup\Si,\qquad
\frac{\p}{\p\nu}\psi_0=0,\quad x\in\om_2.
\end{aligned}\label{6}
\end{equation}
\end{theorem}

\begin{theorem}\label{th2} Suppose the inequalities
\begin{equation*}
\eta_0(\e)\eta(\e)\le a_j(s,\e)\le \eta(\e),\quad
\eta_0(\e)\eta(\e)\le b_j(s,\e)\le \eta(\e),
\end{equation*}
holds, where $\eta_0(\e)$ and $\eta(\e)$ are arbitrary functions
obeying the equalities:
\begin{equation*}
\lim\limits_{\e\to0}\e\ln\eta_0(\e)=0,\quad
\lim\limits_{\e\to0}\frac{1}{\e\ln\eta(\e)}=-A, \quad A=const>0.
\end{equation*}
Then uniform on $\e$ and $\eta$ estimates
\begin{equation}
-c_k\big(\mu(\e)+\e^2+\e\ln\eta_0(\e)\big)\le \l_\e^k-\l_0^k\le
c_k\big(\mu(\e)+\e^2)\label{8}
\end{equation}
hold true. Here $c_k>0$ are constants,
\begin{equation*}
\mu=\mu(\e)=-A-\frac{1}{\e\ln\eta(\e)},
\end{equation*}
$\l_0^k$ are eigenvalues of the problem
\begin{equation}
\begin{aligned}
&-\D\psi_0=\l_0\psi_0,\quad x\in\Om,\qquad \psi_0=0,\quad
x\in\om_1,
\\
& \frac{\p}{\p\nu}\psi_0=0,\quad x\in\om_2, \qquad
\left(\frac{\p}{\p\nu}+A\right)\psi_0=0,\quad x\in\Si.
\end{aligned}\label{5}
\end{equation}
\end{theorem}

\begin{theorem}\label{th3} Suppose the inequalities
\begin{equation*}
0<a_j(s,\e)\le\eta(\e),\quad 0<b_j(s,\e)\le\eta(\e)
\end{equation*}
hold, where $\eta(\e)$ is an arbitrary function obeying an
equality:
\begin{equation*}
\lim\limits_{\e\to0}\frac{1}{\e\ln\eta(\e)}=0.
\end{equation*}
Then uniform on $\e$ and $\eta$ estimates
\begin{equation}
0\le \l_\e^k-\l_0^k\le c_k\mu(\e)\label{9}
\end{equation}
hold true. Here $c_k>0$ are constants,
\begin{equation*}
\mu=\mu(\e)=-\frac{1}{\e\ln\eta(\e)},
\end{equation*}
$\l_0^k$ are eigenvalue of the problem (\ref{5}) as $A=0$.
\end{theorem}

\begin{remark}\label{rm1.2}
We note that the hypotheses of the Theorems~\ref{th1}-\ref{th3}
admit the nonperiodic set $\g_\e$. Moreover, this set may
contain strips of width varying arbitrary, in particular, strips
of fast oscillating width. For instance, the case shown in the
figure is included under consideration.
\end{remark}

\begin{remark}\label{rm1.1}
We stress that the estimates  (\ref{7}), (\ref{8}), (\ref{9})
are best possible. More concretely, under hypothesis of each
theorem there exist function $a_j$ and $b_j$, for those the
degree of convergence has the smallness order exactly as given
in the estimates (\ref{7}), (\ref{8}), (\ref{9}). This statement
will be established in the proofs of
Theorems~\ref{th1}-\ref{th3}.
\end{remark}

\section{Proof of Theorems~\ref{th1}-\ref{th3}}\label{sec:2}

The proof of Theorems~\ref{th1}-\ref{th3} is based  on the
following auxiliary statement.

\begin{lemma}\label{lm1} Suppose the subsets $\g_{\e,*}$ and
$\g_\e^*$ of the lateral surface $\Si$ satisfy
\begin{equation}
\g_{\e,*}\subseteq\g_\e\subseteq\g_\e^*,\label{90}
\end{equation}
and let $\l_{\e,*}^k$ and $\l_\e^{*,k}$ be eigenvalues of the
problem (\ref{1}), (\ref{2}) with $\g_\e$ replaced by
$\g_{\e,*}$ and $\g_\e^*$, respectively. Then the estimates
\begin{equation}
\l_{\e,*}^k\le \l_\e^k\le\l_\e^{*,k}\label{11}
\end{equation}
are valid.
\end{lemma}

This lemma can be proved in a standard way on the base of
minimax property of the eigenvalues of the problem (\ref{1}),
(\ref{2}).

\emph{Proof of Theorem~\ref{th1}.} Let
\begin{equation*}
\g_{\e,*}=\{x: \x\in\p\om, |x_3-\e\pi(j+1/2)|<\e\eta(\e),
j=0,\ldots,N-1\},\quad \g^*_\e=\Si.
\end{equation*}
In accordance with hypothesis, these sets meet the inclusions
(\ref{90}). The set $\g_{\e,*}$ being periodic, the asymptotics
\begin{equation}\label{10}
\l_{\e,*}^k=\l_0^k+\e\l_1^k\ln\sin\eta+o\big(\e(|\ln\eta|+1)\big),
\end{equation}
hold. Here $\l_1^k$ are some constants, $\l_0^k$ are eigenvalues
of the problem (\ref{6}). In the case of simple eigenvalue
$\l_0^k$ this asymptotics was constructed formally in \cite{CR}.
The case of eigenvalue $\l_0^k$ of arbitrary multiplicity was
studied rigorously in \cite{BDU} under assumption $\eta=const$,
$\om$ is a unit circle. These additional assumptions are not
essential and the asymptotics (\ref{10}) can be proved
completely by analogy with \cite{CR,BDU}; the rigorous
two-parametrical estimates for the error term can be easily
established on the base of the results of \cite{VMU}. Since
$\g_\e^*=\Si$, it follows that $\l_\e^{*,k}=\l_0^k$.
Substituting these equalities and asymptotics (\ref{10}) into
(\ref{11}), we get
\begin{equation*}
\l_0^k+c_k\e(|\ln\eta|+1)\le \l_\e^k\le\l_0^k,
\end{equation*}
where $c_k>0$ are some constants. This implies the needed
two-sided estimates. These estimates are best possible. Indeed,
taking $a_j(s,\e)=b_j(s,\e)=\eta(\e)$, we see that $\l_\e^k$
meet asymptotics (\ref{10}), i.e.,
\begin{equation*}
\l_\e^k-\l_0^k=O(\e(|\ln\eta|+1).
\end{equation*}
\qed

\emph{Proof of Theorem~\ref{th2}.} The scheme of the proof is
same with one of Theorem~\ref{th1}. The sets $\g_{\e,*}$ and
$\g^*_\e$ should be defined as follows
\begin{align*}
\g_{\e,*}&=\{x: \x\in\p\om,
|x_3-\e\pi(j+1/2)|<\e\eta_0(\e)\eta(\e), j=0,\ldots,N-1\},
\\
\g_\e^*&=\{x: \x\in\p\om, |x_3-\e\pi(j+1/2)|<\e\eta(\e),
j=0,\ldots,N-1\}.
\end{align*}
By the hypothesis, these sets obey inclusions (\ref{90}). The
corresponding eigenvalues $\l_{\e,*}^k$ and $\l_{\e}^{*,k}$ have
the asymptotics
\begin{align*}
\l_{\e,*}^k&=\l_0^k+\widetilde{\mu}\l_{0,1}^k+\e^2\l_{1,0}^k
+o\big(|\t\mu|+\e^2\big),
\\
\l_\e^{*,k}&=\l_0^k+\mu\l_{0,1}^k+\e^2\l_{1,0}^k
+o\big(|\mu|+\e^2\big).
\end{align*}
Here $\l_{i,j}^k$ are some constants, $\l_0^k$ are eigenvalues
of the problem (\ref{5}),
\begin{equation*}
\t\mu=\t\mu(\e)=-A-\frac{1}{\e\ln\eta_0\eta}.
\end{equation*}
For the case of simple eigenvalue $\l_0^k$ and arbitrary section
$\om$, as well as for the case of multiply eigenvalue  $\l_0^k$
and circular section $\om$ these asymptotics were found in
\cite{CR,BMS}. The technique of these two papers can be easily
applied to the case of arbitrary section $\om$ and multiply
eigenvalue $\l_0^k$. The asymptotics for $\l_{\e,*}^k$ and
$\l_{\e}^{*,k}$, definition of  $\t\mu$ and inequalities
(\ref{11}) yield needed estimates for degree of convergence.
Unimprovability for these estimates is established by analogy
with Theorem~\ref{th1}. \qed

The proof of Theorem~\ref{th3} is similar to proof of
Theorems~\ref{th1},~\ref{th2}. Here sets $\g_{\e,*}$ and
$\g^*_\e$ and asymptotics for corresponding eigenvalues
$\l_{\e,*}^k$, $\l_\e^{*,k}$ (\cite{CR,BMS}) are as follows:
\begin{align*}
&\g_{\e,*}=\emptyset, \quad \g^*_\e=\{x: \x\in\p\om,
|x_3-\e\pi(j+1/2)|<\e\eta(\e), j=0,\ldots,N-1\},
\\
&\l_{\e,*}^k=\l_0^k,\quad
\l_\e^{*,k}=\l_0^k+\l_{0,1}^k\mu+o(\mu),
\end{align*}
where $\l_{0,1}^k$ are some constants, $\l_0^k$ are eigenvalues
of the problem (\ref{5}) with $A=0$.

%and \cite{RefJ}
%\subsection{Subsection title}
%\label{sec:2}
%as required. Don't forget to give each section
%and subsection a unique label (see Sect.~\ref{sec:1}).
%%
%% For one-column wide figures use
%%
%% For two-column wide figures use
%%\begin{figure*}
%%% Use the relevant command for your figure-insertion program
%%% to insert the figure file. See example above.
%%% If not, use
%%\vspace*{5cm}       % Give the correct figure height in cm
%%\caption{Please write your figure caption here}
%%\label{fig:2}       % Give a unique label
%%\end{figure*}
%%\begin{table}
%%\caption{Please write your table caption here}
%%\label{tab:1}       % Give a unique label
%%%\begin{tabular}{lll}
%%\hline\noalign{\smallskip}
%%first & second & third  \\
%%\noalign{\smallskip}\hline\noalign{\smallskip}
%%%number & number & number \\
%%number & number & number \\
%%\noalign{\smallskip}\hline
%%\end{tabular}
%%% Or use
%%\vspace*{5cm}  % with the correct table height
%%\end{table}
%%%
%%% BibTeX users please use
%%% \bibliographystyle{}
%%% \bibliography{}
%%%
%%% Non-BibTeX users please use

%\newpage

\end{document}